\documentclass[12pt,preprint]{aastex}
\begin{document}

\def\wisk#1{\ifmmode{#1}\else{$#1$}\fi}

\def\lt     {\wisk{<}}
\def\gt     {\wisk{>}}
\def\le     {\wisk{_<\atop^=}}
\def\ge     {\wisk{_>\atop^=}}
\def\lsim   {\wisk{_<\atop^{\sim}}}
\def\gsim   {\wisk{_>\atop^{\sim}}}
\def\kms    {\wisk{{\rm ~km~s^{-1}}}}
\def\Lsun   {\wisk{{\rm L_\odot}}}
\def\Zsun   {\wisk{{\rm Z_\odot}}}
\def\Msun   {\wisk{{\rm M_\odot}}}
\def\um     {$\mu$m}
\def\mic     {\mu{\rm m}}
\def\sig    {\wisk{\sigma}}
\def\etal   {{\sl et~al.\ }}
\def\eg     {{\it e.g.\ }}
 \def\ie     {{\it i.e.\ }}
\def\bsl    {\wisk{\backslash}}
\def\by     {\wisk{\times}}
\def\half {\wisk{\frac{1}{2}}}
\def\third {\wisk{\frac{1}{3}}}
\def\nwm2sr {\wisk{\rm nW/m^2/sr\ }}
\def\nw2m4sr {\wisk{\rm nW^2/m^4/sr\ }}
\title{Cosmic infrared background from Population III stars and its effect on spectra of high-$z$ gamma-ray bursts.}

\author{
A. Kashlinsky\altaffilmark{1}}

\altaffiltext{1}{Observational Cosmology Laboratory, Code 665,
Goddard Space Flight Center, Greenbelt MD 20771 and SSAI; E--mail:
kashlinsky@stars.gsfc.nasa.gov}

\begin{abstract}
We discuss the contribution of Population III stars to the near-IR
(NIR) cosmic infrared background (CIB) and its effect on spectra
of high-$z$ high-energy gamma-ray bursts (GRBs) and other sources.
It is shown that if Population III were massive stars, the claimed
NIR CIB excess will be reproduced if only $\sim 4\pm2\%$ of all
baryons went through these stars. Regardless of the precise amount
of the NIR CIB from them, they likely left enough photons to
provide a large optical depth for high-energy photons from distant
GRBs. Observations of such GRBs are expected following the planned
launch of NASA's GLAST mission. Detecting such damping in the
spectra of high-$z$ GRBs will then provide important information
on the emissions from the Population III epoch and location of
this cutoff may serve as an indicator of the GRB's redshift. We
also point out the difficulties of unambiguously detecting the CIB
part originating from Population III in spectra of low $z$
blazars.
\end{abstract}



\section{Introduction}

Zero-metallicity Population III stars (hereafter P3) are thought
to have preceded the normal metal-enriched stellar populations,
but because they would be located at high $z$ they are
inaccessible to direct observations by current telescopes. If
massive, they are expected to have left a significant level of
diffuse radiation shifted today into IR, and it was suggested that
the cosmic infrared background (CIB) contains a significant
contribution from P3 in near-IR, both its mean level and
anisotropies (see review by Kashlinsky, 2005 and references
therein). This has recently received strong support from
measurements of CIB anisotropies in deep Spitzer/IRAC images
\cite{spitzer}. If P3 are responsible for even a fraction of the
claimed NIR CIB they would provide a high comoving density of
photons all the way to the P3 era. In this {\it Letter} we analyze
effects of such photons on spectra of high-$z$ high-energy
gamma-ray bursts (GRBs) and blazars that would be observed with
the upcoming GLAST LAT instrument to 300 GeV. We show that the
entire claimed NIR CIB excess (NIRBE) can be explained if only
$\sim$4\% of the baryons have gone through P3 stars. This would
result in $\sim 0.1 (1+z)^3$ photons/cm$^3$ whose present day
energy is between 1 and 4\um. Such photons would provide a large
optical depth due to photon-photon absorption for GRBs (and other
sources) at energies that will be probed with GLAST. Detecting
this spectral damping in forthcoming GRB observations will provide
an important test of the P3 era parameters.

\section{CIB from Population III}

\begin{figure}[h]
\plotone{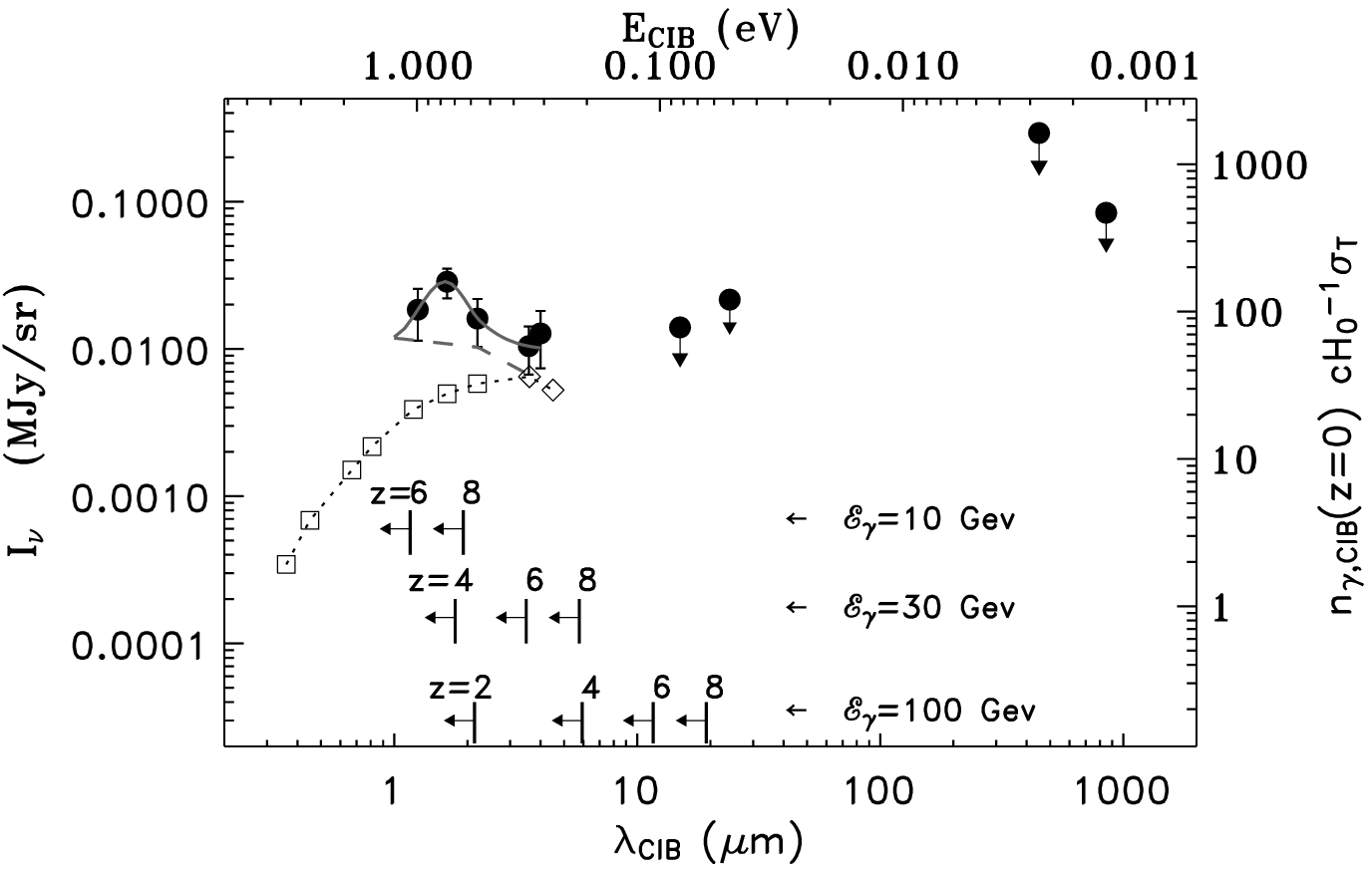} \caption[]{ CIB excess (filled circles) and
``ordinary" galaxy (OG) contributions (open symbols) vs
wavelength. The numbers to compute the CIB excess, i.e. observed
CIB flux minus the OG contribution, are adopted from Fig. 9 of
Kashlinsky (2005 and references therein) and are discussed at
length there (Sec. 5, Table 5 and beyond). Briefly, the net CIB
flux is adopted from Cambresy et al (2001) at 1.25\um, from
Matsumoto et al (2005) at 1.65 \um, from Gorjian et al (2000) and
Matsumoto et al (2005) at 2.2 \um, and from Dwek \& Arendt (1998)
and Wright \& Reese (2000) at 3.5 \um\ and from Matsumoto et al
(2005) at 4 \um. The flux from OG is taken from HST counts out to
2.2 \um\ (open squares from Madau \& Pozzetti 2000) and from
Spitzer/IRAC counts at 3.6 and 4.5 \um\ (open diamonds, Fazio et
al 2004). At $\lambda\gsim$10\um\ no CIB excess was observed and
the levels of CIB are consistent with the net contribution from
OG. The upper limits on the CIB excess there are shown where net
flux from ordinary galaxies is known from SCUBA and ISO
measurements. The CIB level at 450 and 850 \um\ was taken from
Fixsen et al (1998). At 12 and 24 \um\ we adopted the lowest {\it
upper} limits on the net CIB flux using $\gamma$-ray blazar
observations \cite{stanev,renault}. They are also largely
irrelevant for computations of the GRB photons absorption:
vertical bars with left-pointing arrows show the range where
photon-photon absorption is possible for the redshifts and
energies indicated. The thick light-shaded solid and dashed lines
show the CIB excess spectrum used in computing the optical depth
shown in Fig. \ref{tau}. } \label{cib_dc}
\end{figure}

Ay wavelengths $\gsim 10$\um\ the total flux produced by the
observed galaxies matches the levels of the CIB within its
uncertainties, but in the near-IR (NIR) the claimed levels of the
CIB are substantially higher than the net fluxes produced by
galaxies out to flux limits where this contribution saturates (see
review by Kashlinsky 2005 for details). Fig. \ref{cib_dc} shows
the CIB {\it excess} levels (filled circles) over the net flux
from galaxies observed in deep surveys (open symbols); the caption
discusses the details. The excess is significant at $1\mic \la
\lambda \la 4\mic$, the range we term NIR, and its bolometric flux
is \cite{review}:
\begin{equation}
F_{\rm NIRBE}=29 \pm 13 \frac{\rm nW}{\rm m^2sr} \; \; ; \;\;
F_{\rm CIB \; excess} (\lambda\gsim10\micron)\lsim 10 \frac{\rm
nW}{\rm m^2sr} \label{excess}
\end{equation}
At $\lambda \gsim 10$\um\ we evaluated the upper limits shown in
Fig. \ref{cib_dc} described in the caption. The wavelengths
$\gsim$10\um\ contribute little, so we adopt the value of $F_{\rm
NIRBE}$ for what follows.

It was suggested that the NIRBE is produced by massive P3 stars at
high $z$ ($\ga 10$) \cite{santos,salvaterra,ferrara,cooray,kagmm}.
Significant energy release by P3 is suggested from the recent
measurement of CIB anisotropies in deep exposure Spitzer data
\cite{spitzer}. Because P3 stars, if massive, would radiate at the
Eddington limit, where $L\propto M$, the total flux produced by
them is largely model-independent \cite{rees,kagmm}. We reproduce
briefly the argument from Kashlinsky et al (2004): Each star would
produce flux $\frac{L}{4\pi d_L^2}$, where $d_L$ is the luminosity
distance. Because for massive stars $L\propto M$, the total
comoving luminosity density from P3 is $\int n(L) L dL \propto
\Omega_{\rm baryon} f_* \frac{3H_0^2}{8\pi G}$, where $n(L)$ is
their luminosity function and $f_*$ is the mass fraction of
baryons locked in P3 at any given time. In the flat Universe, the
volume per unit solid angle subtended by cosmic time $dt$ is $dV
=c(1+z) d_L^2dt$. Finally, these stars would radiate at efficiency
$\epsilon$ ($\simeq 0.007$ for hydrogen burning). This then leads
to the closed expression for the total bolometric flux from these
objects:
\begin{equation}
F_{\rm bol} = \frac{3}{8\pi} \frac{c^5/G}{4\pi R_H^2}
\langle(1+z)^{-1}\rangle \epsilon f_3 \Omega_{\rm baryon} \simeq
4\times 10^7 \frac{1}{z_3} \epsilon f_3 \Omega_{\rm baryon}h^2
\frac{\rm nW}{\rm m^2 sr}
\end{equation}
Here $f_3$ is the mean mass fraction of baryons locked in P3 stars
and $z_3 \equiv \frac{1}{\langle (1+z)^{-1}\rangle}$ is a suitably
averaged redshift over their era. The total flux is a product of
the maximal luminosity produced by any gravity-bound object,
$c^5/G$, distributed over the surface of the Hubble radius,
$R_H$=$cH_0^{-1}$, and the fairly understood dimensionless
parameters. From WMAP observations we adopt $\Omega_{\rm
baryon}h^2$=0.0224 \cite{bennett} and, since the massive stars are
fully convective, their efficiency is close to that of hydrogen
burning \cite{schaerer}, $\epsilon$=0.007.

Requiring that P3 stars are responsible for the flux given by eq.
\ref{excess} leads to:
\begin{equation}
f_3 = (4.2 \pm 1.9)\times 10^{-3}  z_3 \frac{0.0224}{\Omega_{\rm
baryon}h^2}\frac{0.007}{\epsilon} \label{pop3fraction}
\end{equation}
Assuming $z_3\simeq$10 this is somewhat less than the $\gsim 5\%$
value suggested by Madau \& Silk (2005) and considerably less than
the $\gsim 10\%$ value of Dwek et al (2005). Within the NIRBE
uncertainty, only $\gsim$2\% of the baryons had to go through P3.
This is not unreasonable considering that primordial clouds are
not subject to many of the effects inhibiting star formation at
the present epochs, such as magnetic fields, turbulent heating
etc. The only criterion for P3 formation seems to be that
primordial clouds turning-around out of the primordial density
field have the virial temperature, $T_{\rm vir}$, that can enable
efficient formation of and cooling by molecular hydrogen
\cite{abell,bromm}. Assuming spherical collapse of gaussian
fluctuations and the $\Lambda$CDM model from WMAP observations
\cite{bennett} the fraction of collapsed haloes at $z$=10 with
$T_{\rm vir}\geq$(400,2000)K is $(2.6,5)\times 10^{-2}$ in good
agreement with eq. \ref{pop3fraction} as can be derived from Fig.
2 in Kashlinsky et al (2004).

Eq. \ref{pop3fraction} was evaluated from 1 to 4\um, but with
significant CIB excess flux outside that range, $f_3$ would
increase. However, at wavelengths $\lsim0.1z_3$\um\ the high-$z$
emissions would be below the Lyman break and would be reprocessed
to $\lambda \gsim 1$\um\ (Santos et al 2002). If $z_3<10$, which
is unlikely in light of WMAP polarization data (Kogut et al 2003),
the rest frame Lyman break may be redshifted to $\lsim$1\um, but
the possible extra CIB excess from $<$1\um\ will be compensated by
$f_3$ in eq. \ref{pop3fraction} decreasing with $z_3$. At longer
wavelengths, the CIB excess given by eq. \ref{excess} can at most
increase $f_3$ by $\lsim 30\%$.

The above estimate is subject to two caveats: First, it assumes
the NIR CIB at the levels given by eq. \ref{excess}, which were
derived assuming a specific set of zodiacal light models. The
latter may carry large systematic uncertainties, which are not
included in the formal uncertainties in eq. \ref{pop3fraction}.
Our estimate of $f_3$ is proportional to the CIB excess summarized
in Fig. \ref{cib_dc} and eq. \ref{excess} and would change
accordingly if these are superseded by future measurements.
Secondly, this estimate assumes that P3 were massive stars leading
to the effective efficiency, $\epsilon$, very close to that of the
hydrogen burning, $\epsilon$=0.007. If P3 were less massive than
$\sim 50-100M_\odot$, the effective $\epsilon$ of their energy
release would decrease by a factor of a few \cite{siess} requiring
significantly larger values of $f_3$ for a given $F_{\rm NIRBE}$.
Independently of the CIB considerations, if P3 were less massive
than $\sim 240 M_\odot$, their fraction must be very small in
order not to overproduce the metallicities of the poorest
Population II stars \cite{heger}, leading to CIB fluxes from P3
significantly lower than eq. \ref{excess}.

\section{Optical depth to photon-photon absorption at high
energies at high $z$: application to forthcoming GRB measurements}

If P3 at early epochs produced even a fraction of the claimed
NIRBE, they would supply abundant photons at high $z$. The
present-day value of $I_\nu$=1 MJy/sr corresponds to the comoving
number density of photons per logarithmic energy interval, $d\ln
E$, of $n_\gamma=\frac{4\pi}{c}\frac{I_\nu}{h_{\rm Planck}}$=0.6
cm$^{-3}$ and if these photons come from high $z$ then
$n_\gamma\propto (1+z)^3$. These photons also had higher energies
in the past, $\simeq$(0.1-0.3)$(1+z)$eV, providing an abundance of
absorbers for sufficiently energetic photons at high redshifts via
$\gamma \gamma_{\rm CIB}\rightarrow e^+e^-$ \cite{ahiezer,
nikishov}. Stecker \& de Jager (1993) have pioneered applications
of the $\gamma$-$\gamma$ absorption to constraining the
present-day CIB from high-energy spectra of low-$z$ blazars. Madau
\& Phinney (1996) and Salamon \& Stecker (1998) have considered
effects by evolving normal galaxy populations on potential future
intermediate $z$ ($\gsim 0.5$) blazars. However, as shown in Sec.
2, P3 stars are likely to have provided a far more abundant source
of photons at high $z$ to interact with high-energy gamma-ray
photons.

GRBs are the obvious objects whose high-energy emissions would be
damped by the absorption from the NIRBE photons. This effect was
difficult to detect with EGRET because of its low sensitivity.
Observations by EGRET have detected only 6 GRB's with one of them
being a record energy 18 GeV photon \cite{hurley}. A successor to
the Gamma-Ray Observatory, NASA's GLAST is to be launched in 2007.
Its Large-Area Telescope (LAT) will provide significantly improved
sensitivity needed for detecting high-$z$ GRBs out to ${\cal
E}$=300 GeV with better than 10\% energy resolution above 100 MeV,
and its large field-of-view should detect $\sim$ 100-300 GRBs/yr.
It is expected that $\sim$50 of these would have enough
high-energy photons to measure spectral indices at ${\cal E}>$0.1
GeV with uncertainty better than
0.1\footnote{http://glast.gsfc.nasa.gov/public/resources/pubs/gsd/GSD\_print.pdf}.
We show in this section that the photons produced by the P3 stars
should leave a detectable signature by damping the high-energy
part of the spectra of high-$z$ GRBs.

We denote with $E$ and ${\cal E}$ the present day energies of the
CIB and GRB photons respectively and the primes refer to
rest-frame energies, e.g. $E^{\prime}=E(1+z)$. The photon-photon
absorption, being electromagnetic in nature, has cross-section
$\sim$ that for the Thompson scattering, $\sigma_T$; it is
$\sigma(E^\prime, {\cal E}^\prime, x) =\frac{3}{16}\sigma_T
(1-\beta^2)[2\beta
(\beta^2-2)+(3-\beta^4)\ln(\frac{1+\beta}{1-\beta})]$, where
$\beta = \sqrt{1-\frac{2m_e^2c^4}{E^\prime{\cal E}^\prime(1-x)}}$
and $x=\cos\theta$. The cross-section has a sharp cutoff as
$\beta\rightarrow 1$, peaks at $\simeq \frac{1}{4}\sigma_T$ at
$\beta\simeq 0.7$, and is $\sigma\propto \beta$ for $\beta \lsim
0.6$. The mean free path of GRB photons in the presence of CIB
would be $(n_{\gamma_{CIB}}\sigma)^{-1} \sim 0.8
(\sigma_T/\sigma)(1 {\rm MJy/sr}/I_\nu)(1+z)^{-3}$ Mpc.

The right vertical axis in Fig. \ref{cib_dc} shows the product of
$\sigma_T cH_0^{-1}$ and the comoving photon number density for
given $I_\nu$. The figure also marks the regions defined by the
photon-photon absorption threshold. Discussion in Sec.2 makes it
plausible that, if the NIRBE originates from P3 stars, it must
have a sharp truncation corresponding to the redshifted Lyman
break, or $\simeq 0.1 z_3 \micron$. Thus we assume that, at least
at high $z$, there are few photons at wavelengths that are shifted
today to $<$1\um. At longer wavelengths there is no observational
evidence for CIB excess over that from ``ordinary" galaxies
containing Population I and II, but at the same time only spectra
of the sources of very high energy $\gamma$-rays at very high $z$
would be affected by that range of the CIB. Thus it appears that
there exists a narrow wavelength window of $1\micron \lsim \lambda
\lsim 4\micron$ in which the CIB photons can interact with
high-${\cal E}$, high-$z$ GRBs and probe the emissions from the,
so far putative, P3 era.

We assume a flat Universe dominated by the cosmological constant
and that the NIRBE photons originated from P3 at redshifts higher
than that of the GRBs, so that $n_\gamma \propto (1+z)^3$. In this
case, the optical depth due to photon-photon absorbtion is:
\begin{equation}
\frac{d\tau_{GRB}({\cal E})}{dz} = R_H
\frac{(1+z)}{\sqrt{\Omega_m(1+z)+(1-\Omega_m)(1+z)^{-2}}}
\int_{-1}^1 dx \int_{\frac{2m_e^2c^4}{{\cal E}^\prime(1-x)}}^{
E^\prime_{\rm max}} \sigma(\beta) n_{\gamma_{\rm
CIB}}(\frac{E^\prime}{1+z}) \frac{dE^\prime}{E^\prime}
\end{equation}
where $n_{\gamma_{\rm CIB}}$ is the present-day photon density
corresponding to the observed CIB excess. Fig. \ref{tau} shows the
resultant optical depth and contributions to it from different
$z$. We adopted $E_{\rm max}$=1.24 eV (1 \um) and the form of
$n_{\gamma_{\rm CIB}}$ corresponding to the solid line in Fig.
\ref{cib_dc}.

The onset of $\tau >$1 occurs rapidly at ${\cal E}\gsim
m_e^2c^4/(1 {\rm eV}) (1+z)^{-2} \sim 261 (1+z)^{-2}$ GeV. The GRB
spectra at these energies should either be strongly damped or
there had to be only negligible energy releases from the P3 era.
Given the high values of $\tau$ this would still hold even if P3
era produced only a small part of $F_{\rm NIRBE}$, but assumes
that the latter has a Lyman cutoff redshifted to $\sim$1\um\
today. If the P3 era extended to lower $z$, the Lyman cutoff would
occur at the observer wavelength $\sim 0.1z_3$\um\ and GRB spectra
would be damped at proportionately lower ${\cal E}$. Thus, {\it
the location of this cutoff may also serve as an indicator of the
GRB's redshift}. With the advertised LAT energy resolution of
$<10\%$ at ${\cal E}\gsim$1 GeV, one could determine GRB redshifts
from the damping by the P3 photons to better than $\sim 5\%$.

How robust is this result?  First, photons due to optical counts
of galaxies at much later times do not affect much the GRB spectra
because $\tau \sim z n_\gamma \sigma_T R_H$ at low $z$ and they
contribute $n_\gamma(z=0)\sigma_T R_H\sim$ a few as Fig.
\ref{cib_dc} shows. Furthermore, unlike the P3 photons for which
$n_\gamma\propto(1+z)^3$, those contributed by galaxies would
produce a still smaller contribution at earlier times compared to
P3. Second, even if the NIRBE is not entirely cosmological, Fig.
\ref{tau} shows that P3 emissions should still lead to a very
large optical depth, which scales as $\tau \propto F_{\rm NIRBE}$.
Thus P3 would likely be the dominant contributors to the optical
depth to GRBs at high energies. Finally, the magnitude of $\tau$
at a given bolometric CIB flux should not be sensitive to the CIB
excess spectrum because the (already narrow) range of 1-4 \um\
available to dampen GRBs at high $z$ is further decreased by
$(1+z)$.

Fig. \ref{tau} shows that the optical depth from P3 is very high,
$\tau \gsim 10^2-10^3\left(\frac{F_{\rm NIRBE}}{30 \;{\rm
nW/m^2/sr}}\right)$ which, when combined with the sudden onset of
the optically thick regime, would lead to an identifiable damping
by the P3 photons. The damping will affect progressively lower
energy part of the rest-frame GRB spectra as one moves to higher
$z$. At these energies, the GRB emission is likely produced by the
inverse Compton and the synchrotron self Compton components and is
expected to be high (or even dominate) for the typical Lorentz
factors involved (Piran 2004 and references therein); e.g. Dingus
et al (1998) construct an average spectrum from four EGRET burst
and find the differential photon index of $dn_\gamma/d{\cal E}
\propto {\cal E}^{-(1.95\pm 0.25)}$ out to $\sim$10GeV. It may be
sufficient to use GLAST observations of fairly low $z$ GRBs (say,
$z\sim$2-4 determined spectroscopically in afterglow observations)
to establish the existence of the CIB from P3 epochs; if positive
then the higher $z$ GRBs can be used to further verify that it
comes from the P3 epochs and/or calibrate the $z$ determinations
from the damping.

\section{Is Population 3 detectable with low $z$ blazars?}

\begin{figure}[h]
\plotone{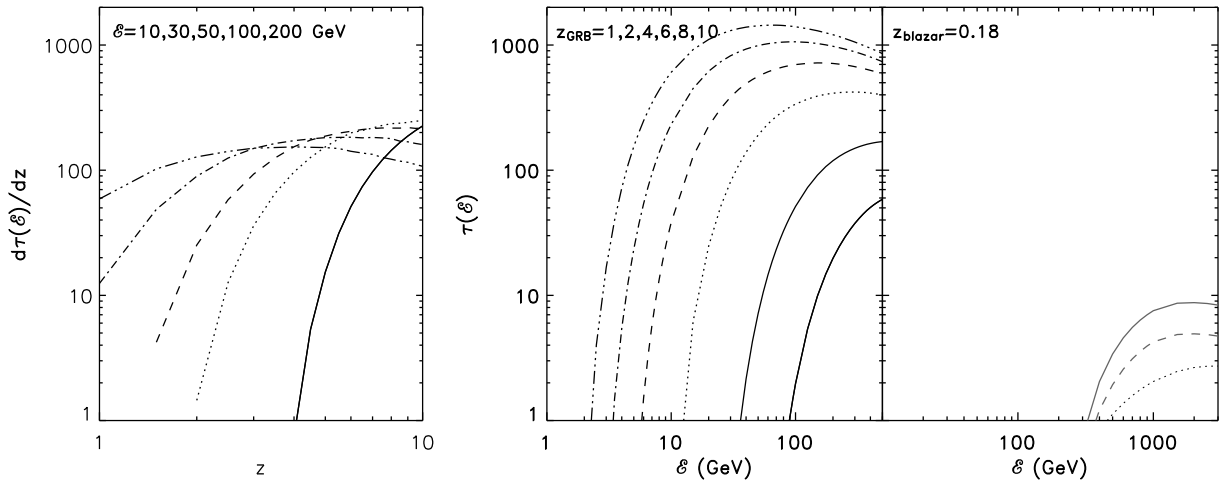} \caption[]{ {\it Left}: $d\tau/dz$ vs $z$ for GRB
photon energies shown in the panel. Solid, dotted, dashed,
dash-dotted and dash-triple-dotted lines correspond to increasing
order in ${\cal E}$. {\it Middle}: The net $\tau$ vs the GRB
photon energy for the GRB redshifts shown in the panel. Solid,
dotted, short-dashed, dash-dotted, dash-triple-dotted  and
long-dashed lines correspond to increasing order in $z$. The range
of redshifts was chosen to avoid overlap between GRBs and P3 era;
the latter is assumed to have ended by $z$=10. Only NIRBE from
Fig. \ref{cib_dc} is assumed in the calculations. This assumption
is fairly safe at larger $z$ as this component gets progressively
larger the ordinary galaxies emissions, but at $z\simeq 1$ the
latter can still contribute \cite{madauphinney}. {\it Right}:
Optical depth to photon-photon absorption for a source at
$z$=0.18. Line notations correspond to Fig. \ref{cib_dc}. Dotted
line assumes only ordinary galaxies measured in deep counts and
that their photons originated at $z\geq0.18$. Thick light-shaded
lines correspond to the NIRBE: solid line assumes both the DIRBE-
and IRTS-based claims at the central points of the measurements
and dashed line assumes the ``minimal" NIRBE with only the
DIRBE-based points (i.e. the 1.65 \um\ point is omitted) and the
CIB levels corresponding to the lower end of the error bars in
Fig. \ref{cib_dc}.} \label{tau}
\end{figure}

Although spectra of more and more distant blazars are now measured
with new instruments, such as HESS
\footnote{http://www.mpi-hd.mpg.de/HESS/}, these blazars are still
too close for unambiguous detection of the P3 emissions. The
farthest blazars with known spectra are at $z\simeq$0.13-0.18 and
the spectra extend to ${\cal E} <$3 TeV (Dwek et al 2005,
Aharonian et al 2005). The right panel in Fig. \ref{tau} shows the
optical depth of a blazar at $z$=0.18 produced by the
$\gamma$-$\gamma$ absorption 1) due to observed galaxies, 2) due
to NIRBE from DIRBE and IRTS, and 3) omitting the IRTS-based point
at 1.65 \um\ and assuming the lower end of the DIRBE-based
results. Even if the entire NIRBE is correct and originates from
P3, the {\it additional} damping from it is small and is more
pronounced only at ${\cal E}>$1 TeV, where measurements and
interpretation are difficult. This is because P3 contribute to the
background over a short range of wavelengths longward of $\sim
1$\um. If P3 were to contribute only a fraction of the NIRBE,
their emissions will not contribute appreciably to the observed
spectra of $z\sim 0.1-0.2$ blazars, but would be seen in the
high-$z$ GRBs. If GLAST collects a large sample of blazars and
other AGNs at $z\gsim 1$, Fig. 2 shows that they can also be used
to probe the emissions from the P3 era.

This work was supported by the NSF under Grant No. AST-0406587. I
thank David Band and Demos Kazanas for useful discussions and
comments on the manuscript.


\clearpage

\clearpage

\end{document}